\documentclass[
    aps,
    prl,
    twocolumn,
    letterpaper,
    10pt,
    floatfix
]{revtex4-1}
\usepackage{bm,enumerate,dcolumn,tikz,graphicx,color,amsmath,amssymb}
\usepackage{epstopdf}

\def\beq{\begin{equation}}
\def\eeq{\end{equation}}
\newcommand{\delete}[1]{}

\newcommand{\be}{\begin{equation}}
\newcommand{\ee}{\end{equation}}

\begin{document}

\title{Sensitivity of atom interferometry to ultralight scalar field dark matter}

\author{Andrew A. Geraci}
\affiliation{Department of Physics, University of Nevada, Reno, Nevada
89557, USA}
\author{Andrei Derevianko}
\affiliation{Department of Physics, University of Nevada, Reno, Nevada
89557, USA}

\begin{abstract}
We discuss the use of atom interferometry as a tool to search for Dark Matter (DM) composed of virialized ultra-light fields (VULFs). Previous work on VULF DM detection using accelerometers has considered the possibility of equivalence principle violating effects whereby gradients in the dark matter field can directly produce relative accelerations between media of differing composition.  In atom interferometers, we find that time-varying phase signals induced by coherent oscillations of DM fields can also arise due to changes in the atom rest mass that can occur between light-pulses throughout the interferometer sequence as well as changes in the Earth's gravitational field.  We estimate that several orders of magnitude of unexplored phase space for VULF DM couplings can be probed due to these new effects.
\end{abstract}
\maketitle

\emph{Introduction--.} Multiple observations in precision cosmology indicate that only
5\% of the total energy density of the universe resides in ordinary
(visible) matter \cite{Bertone2005,Feng2010}, with the rest of the balance coming from dark matter (DM) and dark energy.
So far all the DM signatures
have been purely gravitational and the connection of DM to microscopic physics
remains a mystery.
The main outstanding questions are: the nature of DM constituents and if
they interact with baryonic matter non-gravitationally. Can DM objects be
detected in laboratory-scale experiments? In this paper, we explore the feasibility of using a precision measurement tool,
atomic interferometry \cite{Cronin2009_AtomInterfReview}, for DM searches.

There is a vast range of DM models:
even if DM is composed of elementary particles, the DM particle masses $m_{\rm DM}$ could span a vast 40-order of magnitude mass range, with the lower limit set by the
inverse halo size of smallest galaxies, and the upper limit comes from requiring that these particles do not form black holes. Considering this range of possibilities,
an experimental observation of non-gravitational coupling
is crucial for further progress.
Particle physics experiments (e.g., LUX \cite{Akerib:2013tjd}) search for
DM particles $m_\mathrm{DM}$ that are comparable to the masses of elementary
particles, $\sim 1 -10^{3}$ GeV of the  Standard Model (SM), covering only a narrow sliver of possibilities.

Here, in contrast to particle physics DM searches, we focus on ultralight fields. Recently there were a number of proposals for searches for such fields using precision tools of atomic, molecular, and optical physics.
Among such proposals are magnetometry~\cite{casper,ariadne}, atomic clocks~\cite{DerPos14,ArvHuaTil15}, accelerometers~\cite{StaFla2015}, bar detectors \cite{minabardetector2015}, and laser interferometry~\cite{StaFla2015,laserintf2016}. Depending on the
initial field configuration at early cosmological times,  light fields
could lead to DM  oscillations about the minimum of their potential,
or form stable spatial configurations due
self-interaction potentials. The former
possibility leads to fields oscillating at their Compton frequency and the latter to the formation of topological
defects such as domain walls, strings and monopoles (``topological'' DM \cite{DerPos14}). The properties of  the oscillating virialized ultra-light fields (VULFs~\cite{Derevianko2016a}) have been discussed previously in the context of axions~\cite{Magana2012,Asztalos2006}.  Notice, however,  that axions imply a specific coupling (portal) between the DM and SM particles, while recent proposals considerably widen the classes of  possible portals.

We concentrate on effects of ultralight scalar bosonic oscillating fields.
Such fields, in addition to being the DM candidates, in a certain range of coupling strengths can solve the hierarchy problem \cite{Graham2015}.
We will focus on the mass range $10^{-24}\mathrm{eV}\lesssim
m_{\phi }\lesssim 1 \mathrm{eV}$ for the reasons discussed in~\cite{Derevianko2016a}.  In the standard halo model, during the
galaxy formation, as such particles fall into the gravitational potential,
their velocity distribution in the galactic reference frame becomes
quasi-Maxwellian with the dispersion (virial) velocity $v_{\mathrm{vir}
}\approx 10^{-3}c$. With the dispersion relation $E_{\phi }\approx m_{\phi
}c^{2}+m_{\phi }v_{\phi }^{2}/2$, such fields primarily oscillate at their
Compton frequency $\omega _{\phi }=m_{\phi }c^{2}/\hbar $, ($f_{\phi
}=2.4\times 10^{14}\left( m_{\phi }/ \mathrm{eV}\right) \,\mathrm{Hz}$) although the
spectrum is broadened due to their velocity distribution. The indicated
mass range maps into frequencies \ $10^{-10}\,\mathrm{Hz}\lesssim f_{\phi
}\lesssim 10^{14}\,\mathrm{Hz}$. If the integration time is on the order of
a second, the lower range of the this frequency range would lead to nearly static
effects, while the upper range leads to rapidly oscillating effects on the
experimental timescale.

Further, the number density $n=\rho _{DM}/\left( m_{\phi }c^{2}\right) $ is
given in terms of DM energy density in the Solar system neighborhood $\rho_\mathrm{DM}\approx 0.4 \,\mathrm{GeV/cm^3}$,   in the assumption that the model  saturates the DM energy density.
 The virial velocity
determines the de Broglie wavelength $\lambda _{\phi }^{\mathrm{vir}}=%
(2\pi \hbar)/(m_{\phi }v_{\mathrm{vir}})$. The resulting mode
occupation number $n~\left( \lambda _{\phi }^{\mathrm{vir}}\right) ^{3}\gg 1$ is macroscopic in the indicated mass range
and the bosonic field can be
treated as being classical,
\begin{equation*}
\phi (\boldsymbol{r},t)=\phi _{0}\cos \left( \omega _{\phi }t-\boldsymbol{k}%
_{\phi }\cdot \boldsymbol{r}+\cdots \right) .
\end{equation*}%
Here $\phi _{0}= \hbar/( m_{\phi }c) \,\sqrt{2\rho _{\mathrm{DM}}}$,   the wave vector $k_{\phi }=m_{\phi }v_{\phi }/\hbar $ is distributed
isotropically in the galactic reference frame, $\langle \boldsymbol{k}_{\phi
}\rangle =0$, but in the Earth reference frame moving with respect to the DM
halo with velocity $\boldsymbol{v}_{\oplus }\approx 10^{-3}c,\langle
\boldsymbol{k}_{\phi }\rangle =m_{\phi }\mathbf{v}_{\oplus }/\hbar $.
Otherwise random component of $\boldsymbol{k}_{\phi }$ persists over
coherence lengths $l_{c} = \hbar/ ( m_\phi v_\mathrm{vir})$ or coherence
times $\tau _{c}=1/(\omega_\phi v_\mathrm{vir}^2/c^2)  $ corresponding to $\left( c/v_{%
\mathrm{vir}}\right) ^{2}\sim 10^{6}$ field oscillations~\cite{Derevianko2016a}. For the indicated
mass range, \ $10^{23}\,\mathrm{cm}\gtrsim l_{c}\gtrsim 10^{-1}\,\mathrm{cm}$
and $10^{15}\,\mathrm{s}\gtrsim \tau _{c}\gtrsim 10^{-9} \,\mathrm{s}$.
Notice that the field is coherent over the Earth size if $m_{\phi }\lesssim
10^{-11} \, \mathrm{eV}$, which is consistent with the range of masses that atom interferometry is sensitive to.

As to the DM-SM sector couplings, a systematic approach is
that of the so-called phenomenological portals \cite{Essig:2013lka}, where the gauge-invariant operators of the SM
fields are coupled to the operators that contain DM fields. We focus on the SM-DM interactions in the form of the linear ($n =1$) and quadratic scalar portals ($n =2$)),
\begin{align}
-\mathcal{L}_{n}^\mathrm{int} & =  \left( \sqrt{\hbar c} \phi \right)^{n} \times  \nonumber\\
&\left (\frac{m_{e} \bar{\psi }_{e} \psi _{e}}{%
\Lambda _{n ,e}^{n}} +\frac{m_{p} \bar{\psi }_{p} \psi _{p}}{\Lambda _{n
,p}^{n}} -\frac{1}{4 \Lambda _{n ,\gamma }^{n}} F_{\mu \nu }^{2} +\ldots
\right )\; .  \label{Eq:Portals}
\end{align}
The terms inside the brackets of Eq.~(\ref{Eq:Portals})  are  pieces from the SM sector Lagrangian density. These pieces are weighted with inverses of high-energy scales $\Lambda _{n ,X}$ which parametrize unknown coupling constants.
In particular, $m_{e ,p}$ and $\psi _{e ,p}$ are electron and proton masses and
fields, and $F_{\mu \nu }$ are the electromagnetic filed tensor components.

The main implication of the portals~(\ref{Eq:Portals}) are in modulation of fermion masses and
fundamental constants~\cite{DerPos14}
\begin{equation}
\frac{m_{f}^\mathrm{eff}}{m_{f}}=  1+\frac{\left( \sqrt{\hbar c} \phi(\mathbf{r},t) \right)^{n}}{\Lambda_{n,f}^{n}};\text{\ \ }
\frac{\alpha^\mathrm{eff} }{\alpha} \approx \;  1 +%
\frac{\left( \sqrt{\hbar c} \phi(\mathbf{r},t) \right)^{n}}{\Lambda _{n,\gamma }^{n}} \; .
\label{Eq:TD:VarConst}
\end{equation}
Here $\alpha$ is the fine-structure constant. Similar renormalizations can be written for other SM couplings.




 If the DM field exhibits spatial variations, as in Eq. \ref{Eq:TD:VarConst}, the mass of particles will also acquire gradients, leading to forces on test masses according to the gradient of the field. These forces can in general violate the equivalence principle (EP), and previous work~\cite{dent2008,peteracc} has analyzed the possibility of EP-violating forces due to the spatial gradient of the DM field coupling differently to materials with different constituents. As discussed in Ref. \cite{peteracc}, there are two ways in which this  $-\nabla m c^2$ force can generate a measurable signal in an accelerometer. First, a relative acceleration between two spatially separated test masses can be produced due to a difference in the DM gradient at the location of each mass.  This effect is typically suppressed by the length scale of variations of the gradient $ 1/k_\phi $.  Second, composition-dependent relative accelerations between two test masses can occur even if they are co-located. This effect was discussed in detail in Ref. \cite{peteracc} and we do not reconsider it here, especially due to the fact that the effects of VULF-induced variations in local gravity, as shown below, can be more important.

Indeed, Eq.~(\ref{Eq:TD:VarConst}) also implies that the atomic constituents change their
effective masses due to the DM-SM couplings~(\ref{Eq:Portals}). Thereby we
expect that the total atomic mass would also change $m_{a} \rightarrow
(1 +\left( \sqrt{\hbar c} \phi(\mathbf{r},t) \right)^{n}/\Lambda _{n ,a}^{n}) m_a$, either through the renormalization of elementary constituents masses or coupling constants.  This will lead to the
perturbation of trajectories in light-pulse atom interferometers, as the mass of the recoiling atoms differs for successive laser pulses.
Another, more significant effect comes in the form of variation of the local Earth gravity $g$: when the DM field overlaps with the Earth it makes it effectively heavier/lighter. Then if an atom interferometer is operated as a gravimeter,
it can effectively detect time-varying changes in $g$. In this paper, we develop a theoretical framework for estimating the experimental signals in the form of phase shifts arising in light-pulse atom interferometers due to these effects. We find that several orders of magnitude of unexplored phase space for light DM fields can be probed due to these effects.  A significant improvement is simply due to the fact that in Ref. \cite{peteracc} one measures the gradient of $\phi$ which is independent of $m_\phi$, while our case amounts to measuring the amplitude of $\phi$ which scales as $1/m_\phi$ and gets large at low mass scales.

{\em Change in the Earth gravitational field --} The interaction Lagrangian~(\ref{Eq:Portals}) gives rise to the stress-energy tensor that generates gravitational fields. While we have developed the full-scale formalism based on the Einstein field equation, in the weak-field limit it amounts to modification of the Earth-atom gravitational interaction through the Earth mass renormalization. In particular, it effectively changes gravitational field of the Earth $g$ 
\begin{eqnarray*}
\frac{\Delta g_n}{g} &=& \frac{\Delta M_\oplus}{ M_\oplus} \\ &=&
\left( \frac{  2 \rho_\mathrm{DM} \hbar^3 }{m_\phi^2 c \Lambda_n^2}
\right)^{n/2}  \times \\ &&\frac{1}{2^{(n-1)}}\cos( n \omega_\phi t -n \boldsymbol{k}_{\phi } \cdot
\boldsymbol{r} +\cdots ).
\end{eqnarray*}
For $n=2$, we absorbed the constant part into the conventionally defined $g$.



%

%
%
%

\begin{figure}[!t]
\begin{center}
\includegraphics[width=0.9\columnwidth]{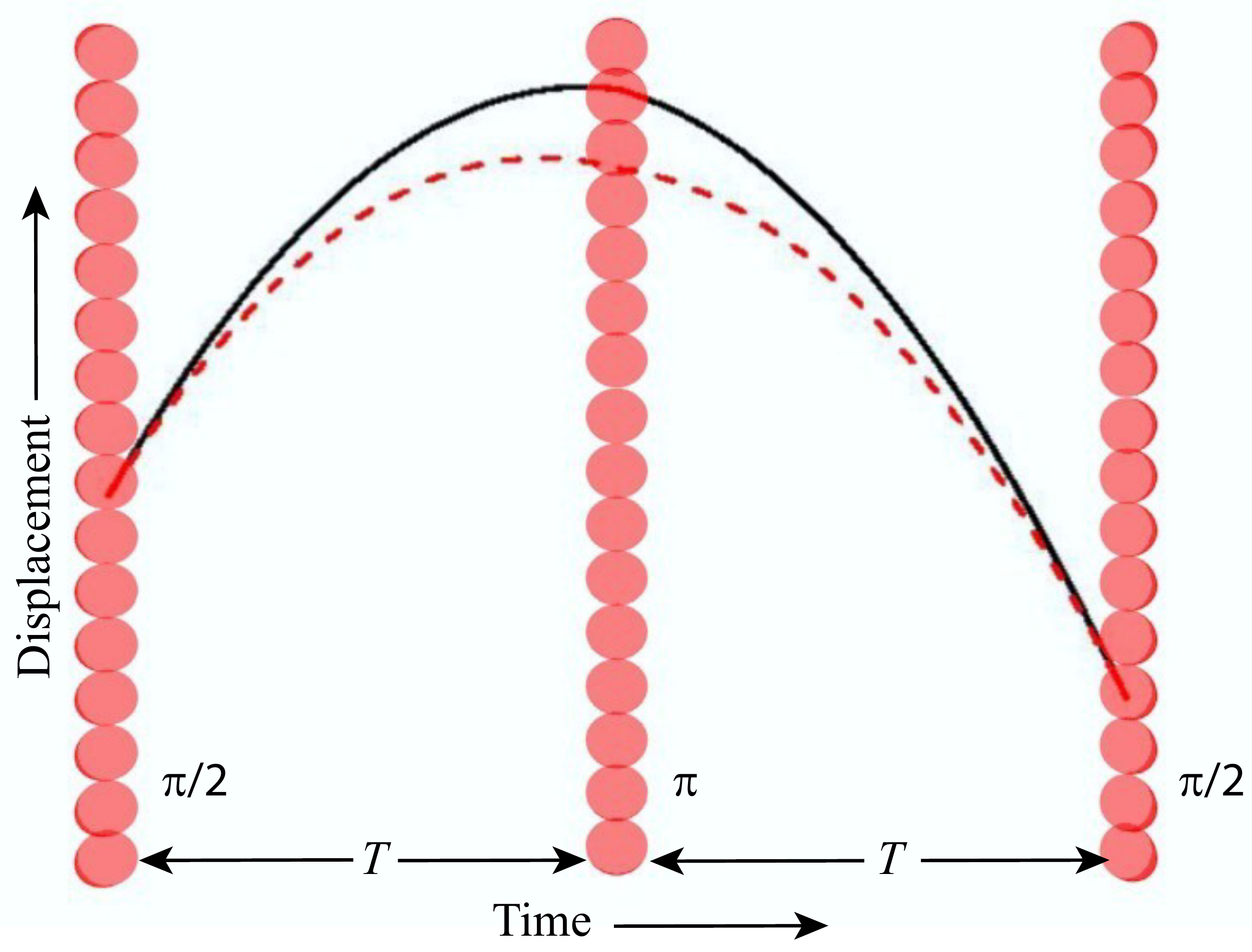}
\caption{Space time diagram for the Mach-Zehnder atom interferometer. Atomic wave packets are split into a superposition state with differing momenta, reflected with a mirror ($\pi$) pulse, and recombined with a final beam-splitter pulse. The final population in state $|1>$ is given by $(1+\cos(\Delta \phi))/2$.
\label{spacetime}}
\end{center}
\end{figure}

\emph{Interferometer signals ---}
To consider a concrete case, we assume the acceleration $g$ due to the Earth and the mass $m$ of atoms in the interferometer are varying sinusoidally in time as \begin{eqnarray} g(t)&=& g_0(1+\delta_g\cos{(\omega t + \theta_0})), \\
 m(t)&=& m_0(1+\delta_m\cos{(\omega t + \theta_0)}) ,\label{massvary}\end{eqnarray} where the amplitude of the fractional change in $g$ and the mass of the atom are denoted as $\delta_g$ and $\delta_m$, respectively, and $\omega=n \omega_\phi$. 
The interferometer sequence we consider is depicted in Fig. \ref{spacetime}. The atoms are launched upwards with a velocity $v_L$. The first $\pi / 2$ pulse splits the atomic wave functions into two trajectories, with $\hbar k_{\rm{eff}}$ of momentum added to the upwards path. This produces a change in the atom's velocity along the upper path equal to the recoil velocity $v_R=\hbar k_{\rm{eff}}/m.$
After time $T$, a $\pi$ pulse is applied, which imparts momentum -$\hbar k_{\rm{eff}}$ to the part of the atomic wave function along the upper path, and momentum +$\hbar k_{\rm{eff}}$ to the component of the atomic wavefunction along the lower path.  After another time $T$ a final $\pi/2$ pulse is applied and the population in the interferometer is read out for atoms along, e.g., the lower path. This population depends on the cosine of the relative phase acquired by the atoms as they have traveled through the interferometer. The total phase shift can be expressed in terms of three contributions, the propagation phase, laser phase, and separation phase: \be \Delta \phi = \Delta \phi_{prop}+\Delta \phi_{laser} + \Delta \phi_{sep}. \ee The propagation phase shift is proportional to the difference in the integral of the classical action along the paths of the interferometer. We compute the laser phase shift as $\Delta \phi_{laser} = k_{\rm{eff}}(z_i-z_{1l}-z_{1u}+z_{2l}),$
where $z_i$ the initial position of the atoms at the time of the first $\pi/2$ pulse, $z_{1l}$ and $z_{1u}$ are the positions of the lower and upper atomic trajectories at the time of the $\pi$ pulse, and $z_{2l}$ is the position of the lower trajectory at the time of the second $\pi/2$ pulse. The separation phase is determined as $\Delta \phi_{sep} = \frac{m}{2\hbar}(v_{2u} - v_R + v_{2l})(z_{2l} - z_{2u}),$
where $v_R$ is the recoil velocity imparted in the final pulse, $v_{2u}$ and $v_{2l}$ are the velocities of the upper and lower trajectories just prior to the final pulse, and $z_{2l}$ and $z_{2u}$ are the positions of the lower and upper trajectories at the time of the final pulse, respectively.

{\it{Effect of VULF DM ---}}
We assume the laser wavelength is kept fixed. (In fact it can change as well but generally by a different fractional amount than the mass of the atom, since it is generally stabilized with respect to a specific atomic transition.) If the atomic mass has changed between the application of the first and second laser pulses, the recoil velocity given to the atoms will be different since
\be
v_R(t)=\hbar k_{\rm{eff}}/m(t)\approx \frac{\hbar k_{\rm{eff}}}{m_0} (1-\delta_m\cos{(\omega t+\theta_0)}).
\label{kick}
 \ee
 Considering the effects of the time variation of the atomic mass and of $g$, the result is the following, kept only to 1st order in $\delta_m$ and $\delta_g$:
\begin{eqnarray*}
\Delta \phi = &-&k_{\rm{eff}} g_0 T^2-\delta_m \frac{2 g_0 k_{\rm{eff}} T}{\omega} (\sin{\omega T}- \sin{2\omega T})\\
&+&\left[\delta_g + 2\delta_m \right] \frac{g_0 k_{\rm{eff}}}{\omega^2}(1-2\cos{\omega T}+\cos{2\omega T})\\ &+& \delta_m \left[\frac{k_{\rm{eff}} (v_L+v_R/2)}{\omega}\right] (2\sin{\omega T}-\sin{2\omega T}).
\end{eqnarray*}
The phase in Eqs. (3) and (\ref{massvary}) at the start of the interferometer sequence is in general unknown. Here we have assumed the initial $\theta_0=0$ for simplicity, to illustrate the amplitude with which the time varying contributions will oscillate. The above expression is modified accordingly for different values of $\theta_0$.  In the low-frequency limit, when the amplitude of the DM field $\phi_0$ becomes large, we can expand taking $\omega T \ll 1$, and find
\begin{equation}
\Delta \phi \approx -k_{\rm{eff}} g_0 (1+\delta_g) T^2 - \delta_m \frac{k_{\rm{eff}}}{\omega} (v_L+v_R/2) (\omega T)^3.
\label{lowfreq}
\end{equation}

Here we do not include the direct acceleration resulting from $\nabla m$ that would generally occur in a EP-violating, composition-dependent way as discussed in previous literature \cite{peteracc}, but only the \emph{indirect} effects from the mass of the atoms and the Earth oscillating with time at frequency $\omega$ that have not been previously considered. In Eq. (\ref{lowfreq}), the terms appearing with $\delta_m$ are due to the time variation in the atomic recoil velocity throughout the interferometer sequence, while the terms appearing with $\delta_g$ result from the variation of the Earth's acceleration $g$.  




To evaluate the phase shift, we take parameters $T=1.34$ s, $g_0=9.8$ m/s$^2$, $m_0=1.44\times 10^{-25}$ kg for $^{87}$Rb, $v_L=10$ m/s and $k_{\rm{eff}}=200 \times 1.6 \times 10^7$ m$^{-1}$, by using large momentum transfer (200-photon recoil) beamsplitters \cite{LMT1,LMT2,LMT3,halfmeter2015}. Assuming $10^{6}$ atoms, with shot-noise limited sensitivity, we can detect a phase of approximately $10^{-3}$ rad/shot. This yields an acceleration sensitivity at the $\delta_g \sim 2\times 10^{-13} /\sqrt{\rm{Hz}}$ level.  In practice laser phase noise and mirror vibrations limit the sensitivity of a single atom interferometer to approximately the n$g/\sqrt{\rm{Hz}}$ level \cite{chu,chinese,biedermann,Zhou2015}.

To attain the ultimate sensitivity a pair of interferometers can be used, in order to cancel out common-mode effects from laser phase fluctuations and platform vibrations \cite{holger_conjugate2009,kasevichphasenoise}. However, this generally suppresses the signal due to the common variations in the time varying acceleration towards the Earth and rest mass of the atoms. By using spatially separated interferometers, the magnitude of the time-varying acceleration from the Earth will produce a detectable time-varying but correlated relative phase shift between the the interferometers, different due to the gradient in the Earth's gravitational field.

To estimate the sensitivity, we consider a setup with two interferometers separated by a baseline of $1$ km which are interrogated with common lasers. We also consider the case where the pair of interferometers is in low-Earth orbit (LEO) vertically separated by $1000$ km. The longer baseline of the space-based approach facilitates a larger difference in the Earth's gravitational field between the two interferometers. Similar arrangements of atomic interferometers have been proposed for gravitational wave searches \cite{kasevichphasenoise,AGIS,AGIS2}, and it is possible such facilities could also be adapted to perform DM searches. In these setups we assume a common isotope, hence the DM differential phase shift results solely from the $\delta_g$ term, while $\delta_m$ makes no contribution. Note any contribution from the difference in the direct DM gradient $\nabla \phi$ at the location of the two interferometers is insignificant by comparison, being suppressed by the length scale of variations of the gradient $1/k_\phi$.

\begin{figure}[!t]
\begin{center}
\includegraphics[width=1.0\columnwidth]{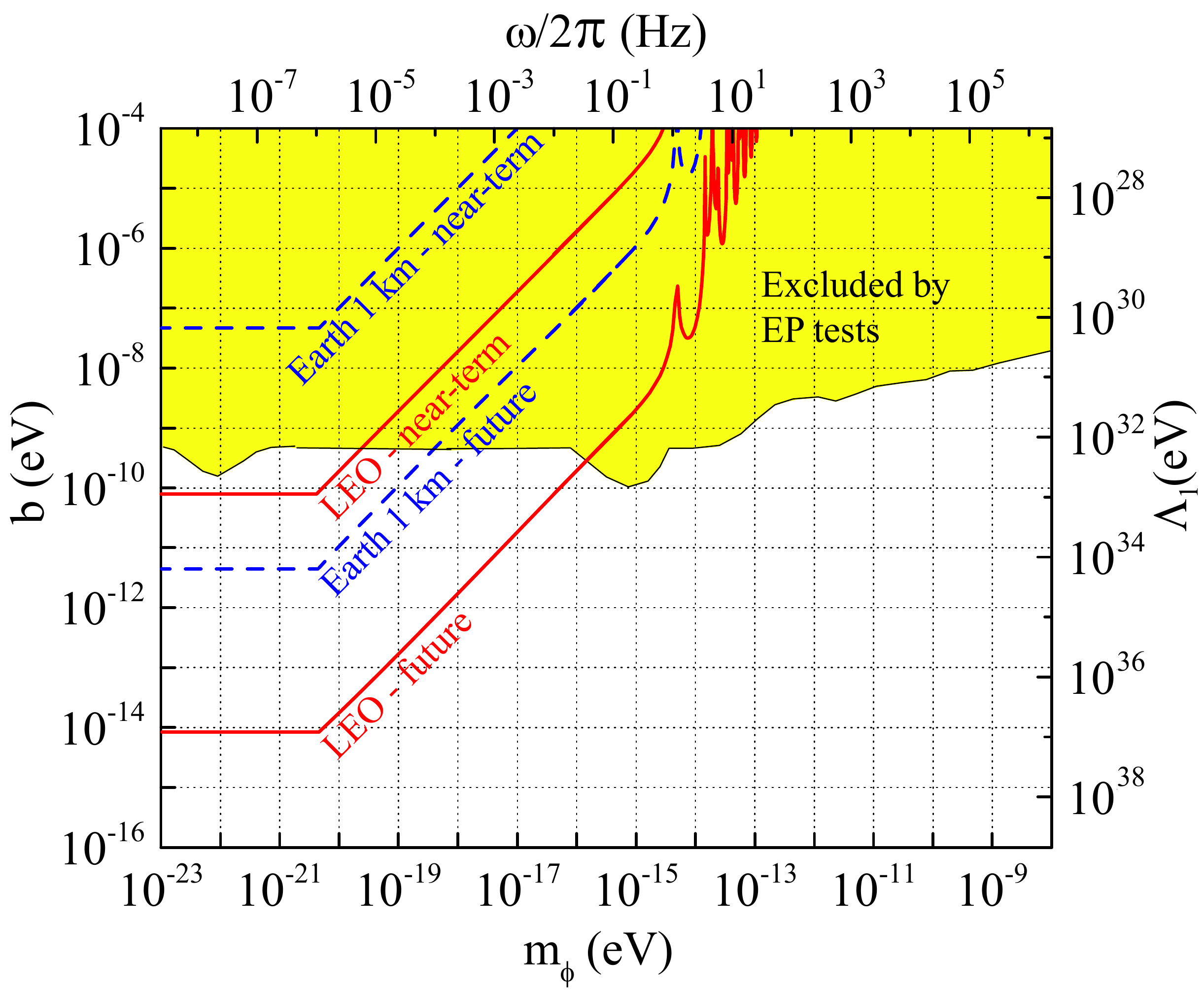}
\caption{Sensitivity in Higgs portal for $n=1$. Previous experimental bounds \cite{wagnerEPV,torsionEPV2,LLR,budkerDMspec} are shown as shaded yellow region, adapted from Ref. \cite{peteracc}. Atom interferometry near-term and future parameters are described in the text.
\label{higgs1}}
\end{center}
\end{figure}

In Fig. \ref{higgs1} we show bounds on the nucleon mass coupling coefficient $\Lambda^1_n$ as defined in Eq. \ref{Eq:TD:VarConst}. Since the nucleon mass is largely determined by $\Lambda_{QCD}$, we can make a connection with previous literature by also showing bounds on the corresponding Higgs portal coupling coefficient $b$ defined in Ref. \cite{peteracc} where $b = 9 m_h^2/ 2 \Lambda_1$.  Here $m_h=125$ GeV/$c^2$ is the Higgs mass. We consider integration over $10^6$ shots, and we assume the DM oscillation is coherent over this time scale, which is reasonable for oscillation frequencies below 1 Hz. We include estimates at the $\delta_g/g \sim 2\times 10^{-13} /\sqrt{\rm{Hz}}$ level of sensitivity (near-term) as well as at the $\delta_g/g \sim 2\times 10^{-17} /\sqrt{\rm{Hz}}$ level (future), which may be possible with larger-momentum-transfer beamsplitters, larger atom number (e.g. $10^8$ atoms), and using an entangled atom source \cite{entangled}. 
We find that several orders of magnitude improvement are possible, with significant improvement beyond the atom-interferometer projections presented in Ref. \cite{peteracc} at low frequencies in particular. This improvement is manifest since the field oscillation amplitude, $\phi _{0}= \sqrt{2\rho _{\mathrm{DM}}} \hbar/( m_{\phi }c)$ becomes very
large for low mass $m_\phi$. The sensitivity does not continue to improve for $\omega/2\pi < 10^{-6}$ Hz since we assume a maximal data set of $10^6$ shots with total interrogation time of order $1$ s. We also find that this approach is competitive with proposed searches based on atomic clocks \cite{ArvHuaTil15}.


\emph{Discussion ---}  In sum, atom interferometry can be a sensitive probe in searches for ultra-light scalar field dark matter through not only direct accelerations of the atoms produced by interactions with dark matter fields, but also through the indirect effects of the inertial and gravitational implications of the variations of the atomic masses and the mass of the Earth. The method shows promise to extend the search for ultra-light scalar field dark matter by several orders of magnitude using the sensitivity of atom interferometers that is realistically achievable in the near-term and farther future. 

\emph{Acknowledgements}
We would like to thank P. Hamilton, H. Muller, D. Schlippert, M. Arvanitaki, S. Dimopoulos, G. Ranjit, E. Rasel, and R. Walsworth for discussions.
This work was supported in part by the US National Science Foundation.

\bibliography{library,library_geraci}

\end{document}